%% file: main.tex
\documentclass[%
 reprint,
 amsmath,amssymb,
 aps,
]{revtex4-2}

\usepackage[utf8]{inputenc}
\usepackage[mode=buildnew]{standalone}
\usepackage{graphicx}% Include figure files
\usepackage{dcolumn}% Align table columns on decimal point
\usepackage{bm}% bold math
\usepackage{braket}
\usepackage{subcaption}
\usepackage{float}
\usepackage{color} %textcolor
\usepackage{amsmath}
\usepackage{amssymb}

\usepackage{xcolor} %Add additional comments

\usepackage[hidelinks]{hyperref}
\def\equationautorefname#1#2\null{%
  Eq.#1(#2\null)%
}
\def\tableautorefname#1#2\null{%
  {table#1#2\null}%
}
\def\sectionautorefname#1#2\null{%
  {Section#1#2\null}%
}
\def\figureautorefname#1#2\null{%
  {Fig.#1#2\null}%
}
\def\subsectionautorefname#1#2\null{%
  {Section#1#2\null}%
}

\newcommand{\change}[1]{\textcolor{black}{#1}}

\begin{document}

\captionsetup{justification=raggedright}

\preprint{APS/123-QED}

\title{Waiting time distributions in hybrid models of motor-bead assays:\\ A concept and tool for inference}% Force line breaks with \\

\author{Benjamin Ertel}
\author{Jann van der Meer}%
\author{Udo Seifert}%
\affiliation{%
 II. Institut für Theoretische Physik, Universität Stuttgart, 70550 Stuttgart, Germany
}%

\date{\today}% It is always \today, today,

\begin{abstract}
In single-molecule experiments, the dynamics of molecular motors are often observed indirectly by measuring the trajectory of an attached bead in a motor-bead assay. In this work, we propose a method to extract the step size and stalling force for a molecular motor without relying on external control parameters. We discuss this method for a generic hybrid model that describes bead and motor via continuous and discrete degrees of freedom, respectively. Our deductions are solely based on the observation of waiting times and transition statistics of the observable bead trajectory. Thus, the method is non-invasive, operationally accessible in experiments and can, in principle, be applied to any model describing the dynamics of molecular motors. We briefly discuss the relation of our results to recent advances in stochastic thermodynamics on inference from observable transitions. Our results are confirmed by extensive numerical simulations for parameters values of an experimentally realized F1-ATPase assay.
\end{abstract}

\maketitle

\section{Introduction}

Molecular motor proteins convert chemical input energy into mechanical work and are therefore one key constituent of living systems. As they operate far from equilibrium but still at a well-defined temperature, their description within the framework of stochastic thermodynamics is well-founded \cite{qian1997,andrieaux2006,gaspard2007,seifert2012,chowdhury2013}. In experiments, a direct observation of the dynamics of a single motor is challenging due to its small size. One possibility to circumvent this problem is attaching a significantly larger bead to the motor. For the resulting motor-bead assay, the motion of the motor can be reconstructed from the dynamics of the bead \cite{ritort2006,herbert2008,veigel2011,ariga2018,bustamante2022}. 

To describe motor-bead assays in and out of equilibrium, various qualitatively different models have been proposed. Discrete Markov network models emphasize the changes of the biochemical configuration of the motor protein \cite{kolomeisky2007,liepelt2007,lipowsky2008,astumian2010,kolomeisky_book}, whereas continuous models based on an overdamped Langevin equation focus on the observable diffusive dynamics of the bead \cite{juelicher1997,reimann2002,aithaddou2003,astumian2016}. Hybrid models include the dynamics of both assay constituents by coupling the discrete dynamics of the motor to the continuous dynamics of the bead via an effective potential \cite{xing2005,zimmermann2012,zimmermann2015,gupta2018,blackwell2019,brown2020,gupta2022,leighton2022}.

Despite the theoretical significance of the aforementioned model classes, their direct applicability to experimental data is limited from an operational point of view. In reality, the dynamics of motor-bead assays is only partially accessible, as only trajectories of the bead are observed in experiments. Since the effective dynamics generating these bead trajectories is non-Markovian due to the coupling, naive overdamped Langevin models or Markov random walk models of the bead cannot reveal underlying mechanisms of the motor \cite{wang2008,brown2019,berezhkovskii2020,godec2023}. 

Constructing appropriate effective models for both motor and bead prove challenging, because even if the model is constructed to mimic the effective bead dynamics correctly, characteristic properties of the motor remain inaccessible \cite{godec2023}. Beyond merely fitting the dynamics, the concept of thermodynamic inference aims at deducing intrinsic properties of a partially accessible system by combining the observable statistics with thermodynamic consistency conditions \cite{barato2014,gingrich2016,pietzonka2016,seifert2018,seifert2019,horowitz2020}. Recent advances place emphasis on waiting time distributions and inter-transition statistics \cite{berezhkovskii2006, PRX,harunari2022, PRL}, which, as discussed in Ref. \cite{berezhkovskii2019}, provide a promising starting point for motor-bead assays too. For example, an experimentally observed broken time-reversal symmetry of transition waiting time distribution\change{s} out of equilibrium \cite{gladrow2019} can rule out low-dimensional models that are too limited to account for the presence of hidden cycles \cite{berezhkovskii2021}.

In this work, we make use of waiting time distributions to infer characteristic, hidden properties of the molecular motor of a motor-bead assay. We describe the assay with a hybrid model, which couples the continuous movement of the bead to discrete steps of the motor. Assuming that only the trajectory of the bead can be observed, we analyze how identifying waiting times between particular transition events allows us to infer the step size of the motor merely from trajectory data of the bead. Knowing the step size, we further demonstrate how the driving affinity of a full motor cycle can be identified, which also provides a non-invasive method to recover the stalling force. We illustrate our findings with simulations of the model for parameter values corresponding to the experimentally realized F1-ATPase assay from Ref. \cite{toyabe2010}. To demonstrate the versatility of our method, we additionally show its applicability to the model for parameter values that lead to strong fluctuations in the bead trajectory thus blurring any visible steps.

The paper is structured as follows. In Section~\ref{Sec:Principles}, we introduce the hybrid model for the motor-bead assay and make the necessary steps to introduce inference tools for the bead trajectory data. This discussion also includes the coarse-graining procedure used to identify transitions of the bead and the definition of appropriate waiting time distributions. We apply the derived inference procedure in Section~\ref{Sec:Hybrid} to recover motor characteristics from simulations of the hybrid model for the experimentally realized F1-ATPase assay. In Section~\ref{Sec:Discussion}, we discuss the role of the waiting time distributions, with emphasis on the relation to similar results for partially accessible Markov networks from stochastic thermodynamics. We conclude in Section~\ref{Sec:Conclusion} by outlining future directions and perspectives.

\section{Setup and theory}\label{Sec:Principles}

\subsection{Hybrid motor-bead model}

\begin{figure}[bt]
    \begin{subfigure}[t]{0.45\textwidth}
      \centering
        \includegraphics[width=\linewidth]{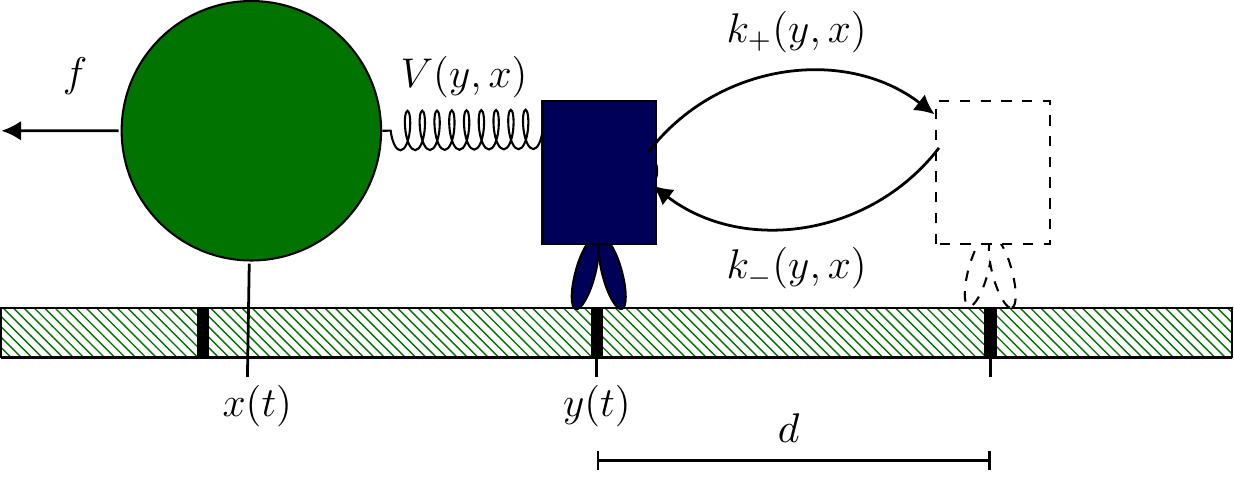}
    \end{subfigure}
	\caption[Motor Draw]{Illustration of the hybrid motor-bead model. The motor at position \change{$y(t)$} steps with step size $d$ along a one-dimensional track in forward or backward direction with corresponding transition rates \change{$k_\pm(y,x)$}. The bead at position $x(t)$ is coupled to the motor via the potential \change{$V(y,x)$} and is subject to an external force $f$.}
	\label{Fig:Mot_Draw}
\end{figure}

On a fundamental level of description, a motor-bead assay consists of a motor protein linked to a bead via an effective potential. We start by introducing a minimal hybrid model for the dynamics \cite{zimmermann2012}. As sketched in Figure~\ref{Fig:Mot_Draw}, at time $t$, the motor and bead position are \change{$y(t)$} and $x(t)$, respectively. The motor moves between discrete positions along a one-dimensional track by making steps of size $d$ from \change{$y_i$} to \change{$y_{i+1} = y_i\pm d$}. Setting $k_{B}T = 1$, we associate each step to a chemical reaction with corresponding free energy change $\Delta\mu$ in the solvent, e.g.,
\begin{equation}
    \Delta\mu = \mu_{ATP} - \mu_{ADP} - \mu_{P}
    \label{Eq:DelMu}
\end{equation}
for a motor driven by ATP hydrolysis. Furthermore, we assume a dilute solution which implies that the chemical potential $\mu_i$ of molecule $i$ is determined by the corresponding concentration $c_i$ \cite{gaspard2007}.

This motor dynamics correspond to an asymmetric random walk in continuous time between discrete states with transition rates $k_{\pm}$. We model the coupling of strength $\kappa$ between motor and bead by the effective harmonic potential
\begin{equation}
    \change{V(y,x)} = \frac{\kappa}{2}(\change{y(t)}-x(t))^2
    ,\label{Eq:V}
\end{equation}
which is illustrated in Figure~\ref{Fig:Mot_Draw} by a spring whose rest length formally is equal to $0$. Following mass action law kinetics and Kramers theory \cite{zimmermann2012,fisher1999}, this coupling affects the transition rates according to
\begin{equation}
  \change{k_\pm(y,x)} = w_{0} \exp\left[\Delta\mu_{\pm} - \change{V(y\pm d\Theta_{\pm},x)} + \change{V(y,x)}\right]
\label{Eq:k}     
.\end{equation}
Here, $\Delta\mu_{\pm}$ corresponds to the chemical free energy change of a forward or backward transition, $w_{0}$ is the motor specific attempt frequency and $\Theta_{+}$ and $\Theta_{-}$ are the motor specific load sharing factors. As discussed in Ref. \cite{zimmermann2012}, $\kappa$, $w_{0}$, $\Theta_{+}$ and $\Theta_{-}$ can be deduced for a given motor by a combination of theoretical considerations and fitting of the mean local velocity.

In contrast to the motion of the motor, the bead moves continuously. Therefore, we describe the corresponding dynamics by an appropriate overdamped Langevin equation,
\begin{equation}
    \Dot{x}(t) = (1/\gamma)\left[\change{-}\partial_x \change{V(y,x)} - f\right] + \zeta(t)
    ,\label{Eq:Langevin}
\end{equation}
which includes the force generated by the motor via \change{$V(y,x)$} and an externally applied force $f$. The effective friction coefficient $\gamma$ models the influence of the solvent and the size and shape of the bead \cite{gaspard2007,kumiko2010,zimmermann2012}, whereas $\zeta(t)$ is the random force modeling thermal fluctuations with $\langle\zeta(t)\rangle = 0$ and $\langle\zeta(t_1)\zeta(t_2)\rangle = (2/\gamma)\delta(t_1 - t_2)$. The driving affinity of the full assay is given by $\Delta\mu - fd$, which incorporates the motion of motor and bead.

\begin{figure*}[bt]
    \begin{subfigure}[t]{0.32\textwidth}
      \centering
        \includegraphics[width=\linewidth]{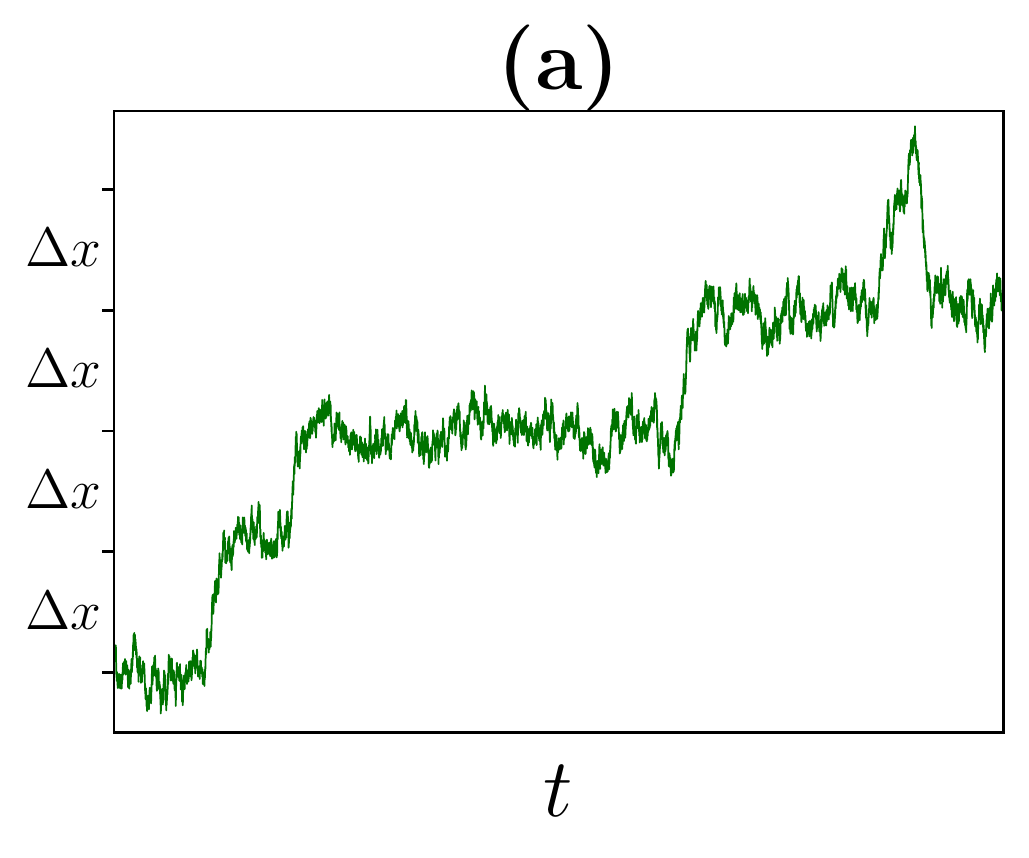}
    \end{subfigure}
    \hfill
    \begin{subfigure}[t]{0.32\textwidth}
      \vspace{-4.785cm}
      \centering
        \includegraphics[width=\linewidth]{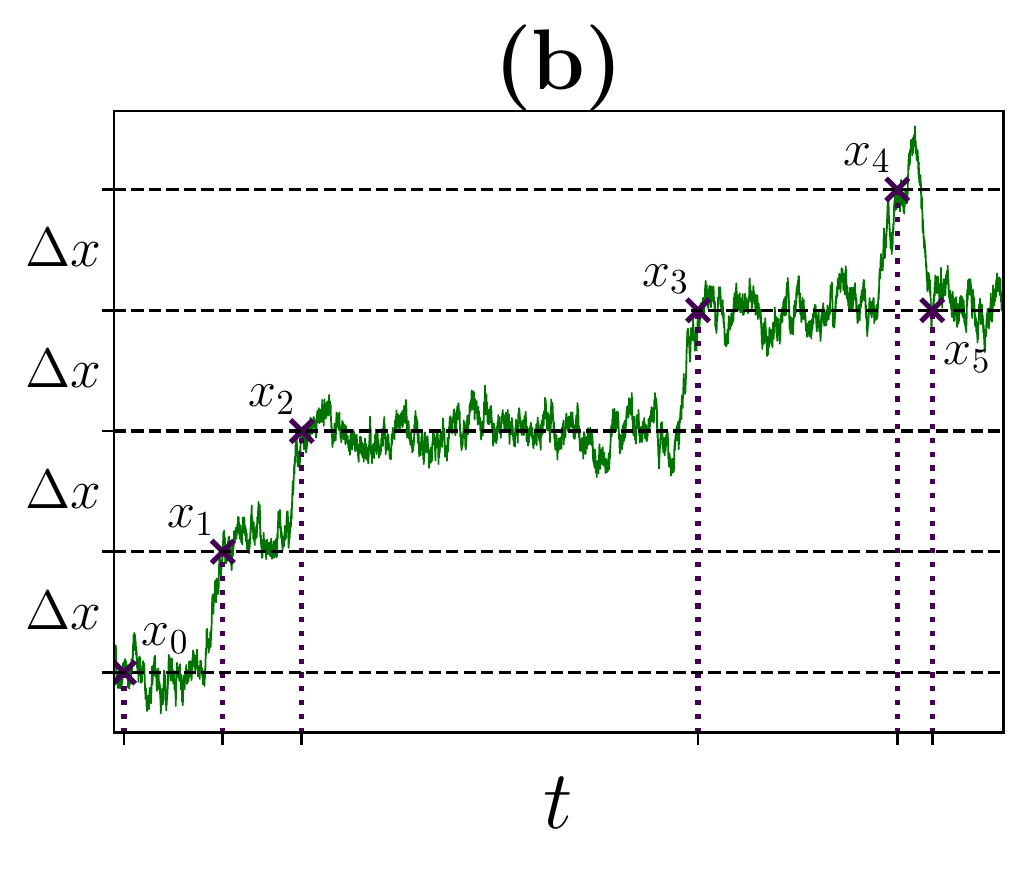}
    \end{subfigure}
    \hfill
    \begin{subfigure}[t]{0.32\textwidth}
      \vspace{-4.78cm}
      \centering
        \includegraphics[width=\linewidth]{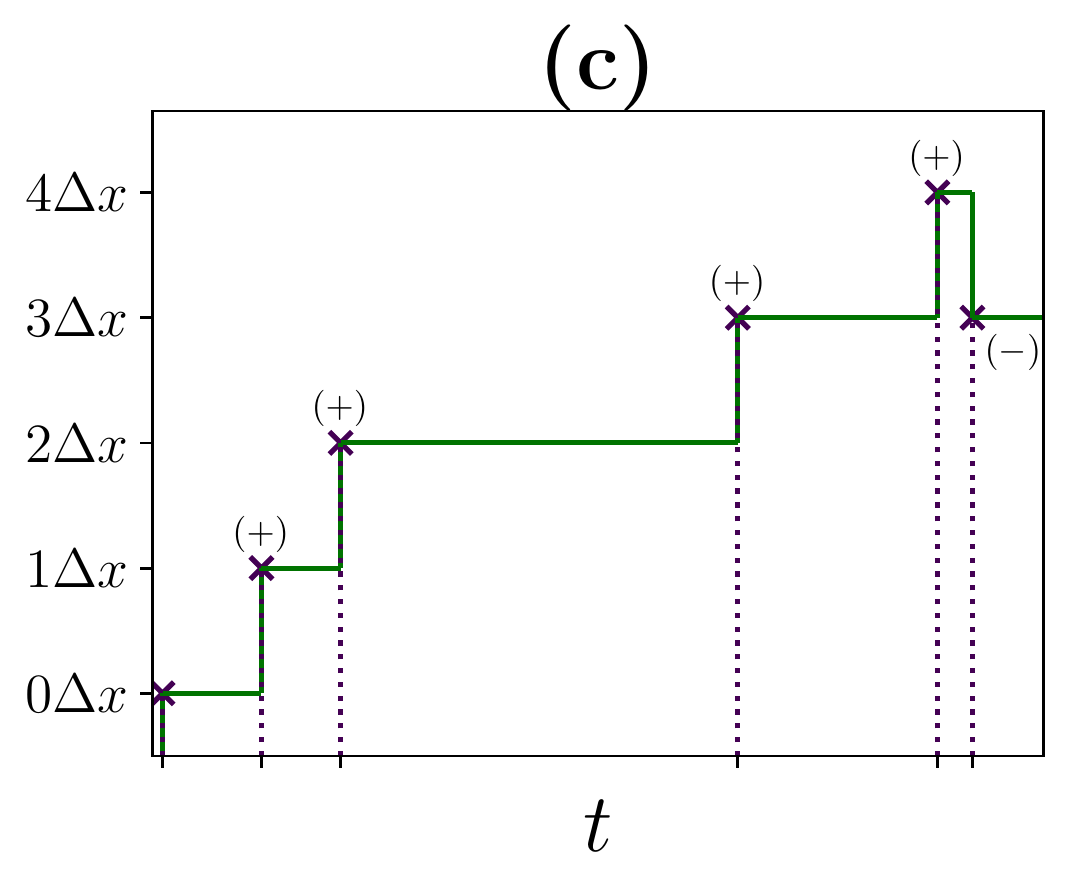}
    \end{subfigure}
    \caption[Inference Milestoning]{Illustration of the coarse-graining procedure for an exemplary bead trajectory. (a) Choice of the reference spacing. Starting from an observed bead trajectory, we choose a value for $\Delta x$. (b) Construction of the effective description. For the chosen $\Delta x$, we identify crossing events $x_i$ along the trajectory. Starting from $x_{i-1}$, a crossing event $x_{i}$ corresponds to positive or negative coverage of $\Delta x$ at a subsequent point later in time. (c) Alternative transition-based interpretation. By interpreting positive coverage of $\Delta x$ as a $+$-transition and negative coverage of $\Delta x$ as a $-$-transition, transitions can be identified for the observed bead trajectory.}
\label{Fig:Milestones}
\end{figure*}

\subsection{Identifying transitions via milestoning}

Suppose we record the exemplary bead trajectory $x(t)$ of a motor-bead assay shown in Figure~\ref{Fig:Milestones} (a). After the start from $x(0) = x_0$, $x(t)$ will eventually cover a distance $\Delta x$ for the first time, i.e., it crosses the point $x_1$, which is either $x_0 + \Delta x$ or $x_0 - \Delta x$. Note that for now, the spacing $\Delta x$ is an arbitrary parameter. Subsequently, $x(t)$ will cover the distance $\Delta x$ again at a later point in time by crossing the point $x_2 = x_1 + \Delta x$ or $x_2 = x_1 - \Delta x$. By repeating this partition until reaching the endpoint of the trajectory, we obtain a coarse-grained description in terms of crossing events $x_i$ that satisfies $x_{i} = x_{i-1} \pm \Delta x$ for all $i=1,...,M$, with $x_0 = x(0)$. This procedure is illustrated in Figure~\ref{Fig:Milestones} (b).

From the perspective of an external observer, a crossing event $x_i$ of the bead represents a completed forward or backward transition of length $\pm \Delta x$. In the following, we refer to these transitions of the bead via their sign, e.g., by denoting a single forward transition as $+$ or two subsequent backward transitions as $--$. The alternative transition-based interpretation of the coarse-grained bead trajectory in Figure~\ref{Fig:Milestones} (b) is illustrated in Figure~\ref{Fig:Milestones} (c). If $\Delta x$ is chosen small, random fluctuations will dominate the statistics, which is evident from inspecting Figure~\ref{Fig:Milestones} (a) and (b).

Conceptually, the applied coarse-graining procedure is equivalent to the method of milestoning \cite{faradjian2004,schuette2011,elber2020,hartich2021}. Covering the distance $\pm\Delta x$ corresponds to a crossing event $x_i$, which can be interpreted as passing a milestone. After reaching this milestone, the coarse-grained state of the system, i.e., the position of the bead, is updated. Thus, this type of event-based coarse-graining retains precise information about the system at particular times solely based on observable statistics, without introducing artificial descriptions as for example effective compound states \cite{rahav2007,pigolotti2008,puglisi2010,knoch2015,seiferth2020}.    

\subsection{Transition statistics and conditioned counting}

In the coarse-grained description, a trajectory is characterized by the sequence of subsequent forward and backward transitions and their in-between waiting times. By counting, we obtain the number of forward and backward transitions along the trajectory, which we denote as $n_{\Delta x}^+$ and $n_{\Delta x}^-$, respectively. Both quantities can be defined as
\begin{equation}
    n_{\Delta x}^\pm \equiv \sum_{m=1}^{M} \delta_{x_m, x_{m-1}\pm\Delta x},
    \label{Eq:Count_+}
\end{equation}
for a trajectory consisting of $M$ crossing events because each crossing event is either a forward transition or a backward transition. For the trajectory shown in Figure~\ref{Fig:Milestones} (c), $n_{\Delta x}^+$ equals four and $n_{\Delta x}^-$ equals one.

If the observed bead trajectory describes a free Brownian particle in some potential landscape, knowing its position determines the state of the system completely. However, in the case of motor-bead assays, the bead is coupled to a hidden degree of freedom, the motor, which introduces memory effects in the bead dynamics. We account for these memory effects by collecting statistics for doublets of transitions. The corresponding counting observables can be defined similar to Equation~\ref{Eq:Count_+} via
\begin{equation}
     n_{\Delta x}^{++} \equiv \sum_{m=2}^{M} \delta_{x_m, x_{m-1}+\Delta x} \delta_{x_{m-1}, x_{m-2}+\Delta x},
   \label{Eq:Count_++} 
\end{equation}
for the total number of doublets of subsequent forward transitions $n_{\Delta x}^{++}$, with $n_{\Delta x}^{-+}$, $n_{\Delta x}^{+-}$ and $n_{\Delta x}^{--}$ defined accordingly. Neglecting the first transition, we have  
\begin{align}
    n_{\Delta x}^+ & = n_{\Delta x}^{++} + n_{\Delta x}^{-+}
    \label{Eq:n_+_Con} \\
    n_{\Delta x}^- & = n_{\Delta x}^{--} + n_{\Delta x}^{+-}
    \label{Eq:n_-_Con}
,\end{align}
since any transition is preceded by a previous forward or backward transition. For the trajectory shown in Figure~\ref{Fig:Milestones} (c), $n_{\Delta x}^{++}$ equals three, $n_{\Delta x}^{+-}$ equals one and $n_{\Delta x}^{-+}$ and $n_{\Delta x}^{--}$ are both equal to zero. 

To extract transition statistics of observed bead trajectories, we introduce conditioned transition probabilities based on the defined counting observables. The probability for a $\pm$-transition following a transition of the same type is given by
\begin{align}
    p(+|+) & = \frac{n_{\Delta x}^{++}}{n_{\Delta x}^{++}+n_{\Delta x}^{-+}}
    \label{Eq:CC_PP} \\
    p(-|-) & = \frac{n_{\Delta x}^{--}}{n_{\Delta x}^{--}+n_{\Delta x}^{+-}}
    \label{Eq:CC_MM}
.\end{align}
We emphasize that conditioning on the previous transition contains additional information only because the observed effective dynamics is non-Markovian. For the same reason, resolving the waiting times between individual transitions includes additional information. We include this information in our statistics as waiting time distributions $\psi_{\pm \to \pm}^{\Delta x}(t)$, which form the time-resolved analogues of the corresponding transition probabilities in Equation~\ref{Eq:CC_PP} and Equation~\ref{Eq:CC_MM} leading to
\begin{eqnarray}
    &&\psi_{\pm \to \pm}(t)\equiv p(x_{m+1}-x_{m}=\pm\Delta x;\nonumber\\ &&T_{m+1}-T_{m}=t|x_{m}-x_{m-1}=\pm\Delta x)
    \label{Eq:CC_Psi}
,\end{eqnarray}
where $T_{i}$ is the time of the $i$-th crossing event $x_{i}$. In other words, Equation~\ref{Eq:CC_Psi} is the probability density that the crossing event $x_{m+1}$ corresponds to a $\pm$-transition and is measured at time $T_{m+1} = T_{m} + t$ given that the previous crossing event $x_{m}$ corresponds to a $\pm$-transition and was measured at time $T_{m}$. To evaluate statistical data, we use a finite $\Delta t$ to obtain a histogram of the continuous waiting time distribution.

As inference quantities, we consider ratios of waiting time distributions of the form 
\begin{equation}
    a_{\Delta x}(t) \equiv \ln \frac{\psi^{\Delta x}_{+ \to +}(t)}{\psi^{\Delta x}_{- \to -}(t)}
    \label{Eq:a_deltax_t}
,\end{equation}
which allow us to highlight differences between $\psi^{\Delta x}_{+ \to +}(t)$ and $\psi^{\Delta x}_{- \to -}(t)$. In a similar fashion, we define the time-independent analog of $a_{\Delta x}(t)$ as
\begin{equation}
    a_{\Delta x} \equiv \ln \frac{p(+|+)}{p(-|-)}
    ,\label{Eq:a_deltax}
\end{equation}
with $p(+|+)$ and $p(-|-)$ defined in Equation~\ref{Eq:CC_PP} and Equation~\ref{Eq:CC_MM}, respectively. Equation~\ref{Eq:a_deltax_t} and Equation~\ref{Eq:a_deltax} define the crucial quantities of this work, which remain without justification for now. Their power as a tool for inference and their thermodynamic interpretation will be discussed in the next two chapters.

\begin{figure*}[bt]
    \begin{subfigure}[t]{0.33\textwidth}
      \centering
        \includegraphics[width=\linewidth]{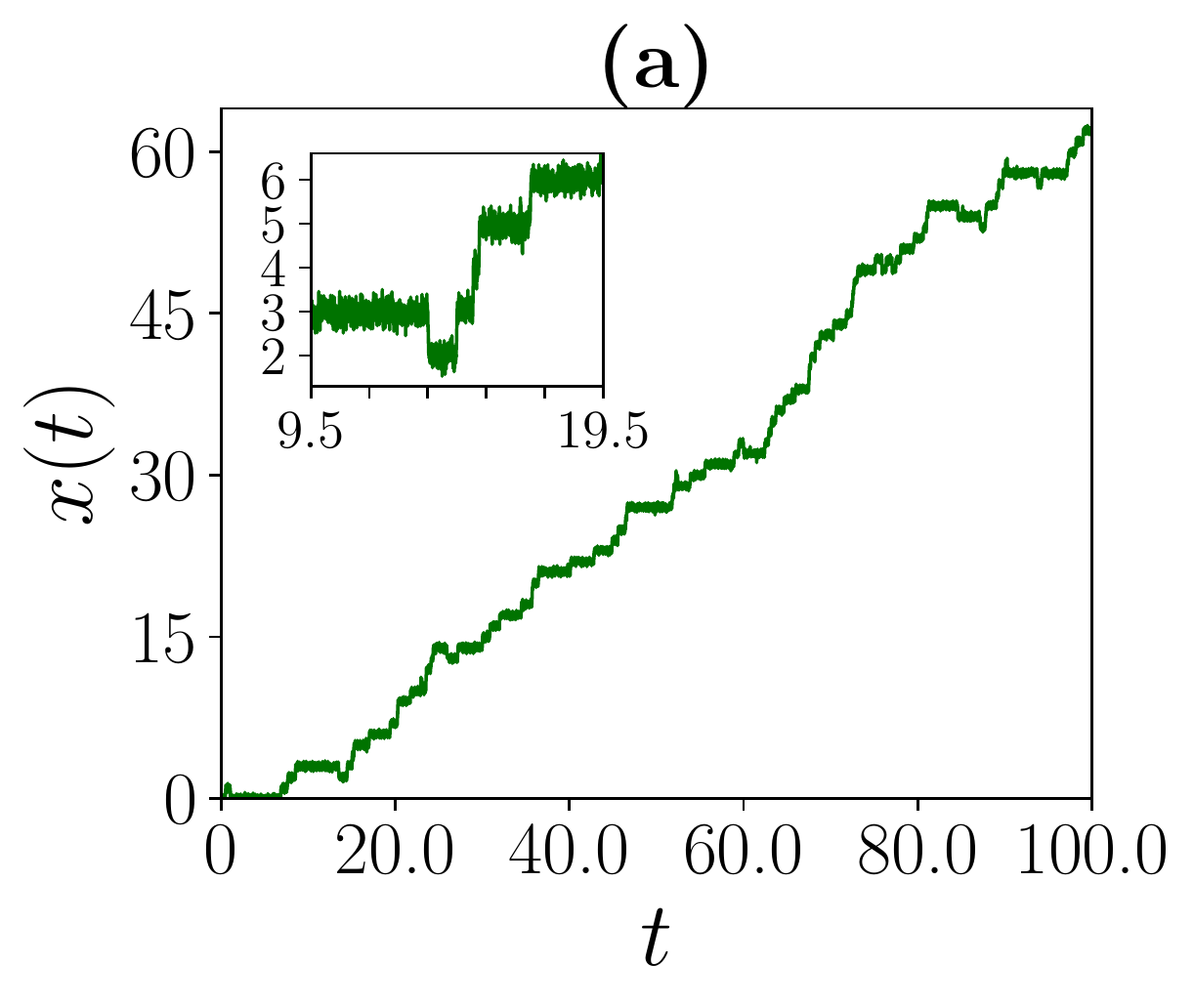}
    \end{subfigure}
    \hfill
    \begin{subfigure}[t]{0.325\textwidth}
      \centering
        \includegraphics[width=\linewidth]{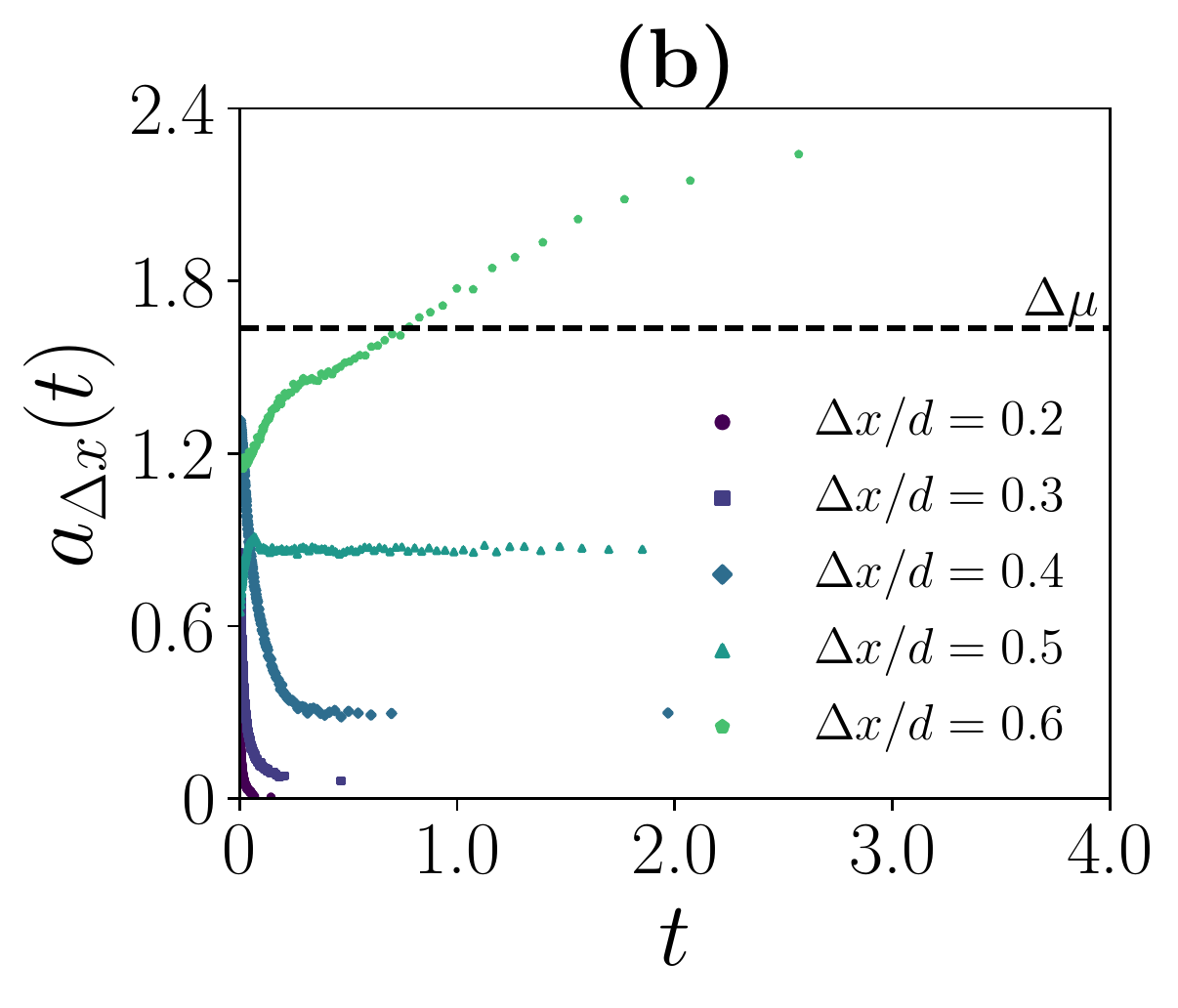}
    \end{subfigure}
    \hfill
    \begin{subfigure}[t]{0.325\textwidth}
      \centering
        \includegraphics[width=\linewidth]{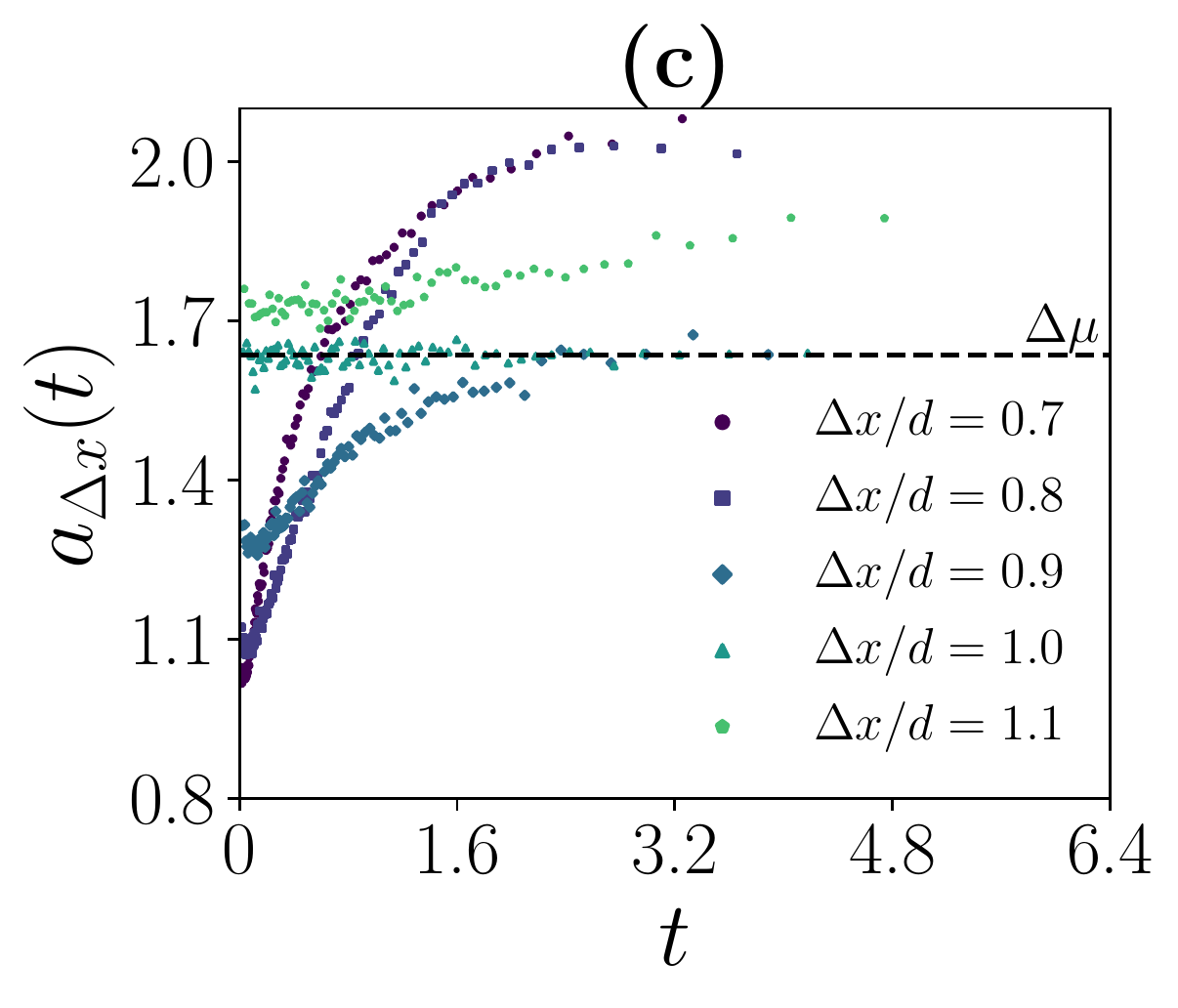}
    \end{subfigure}
    \begin{subfigure}[t]{0.332\textwidth}
      \centering
        \includegraphics[width=\linewidth]{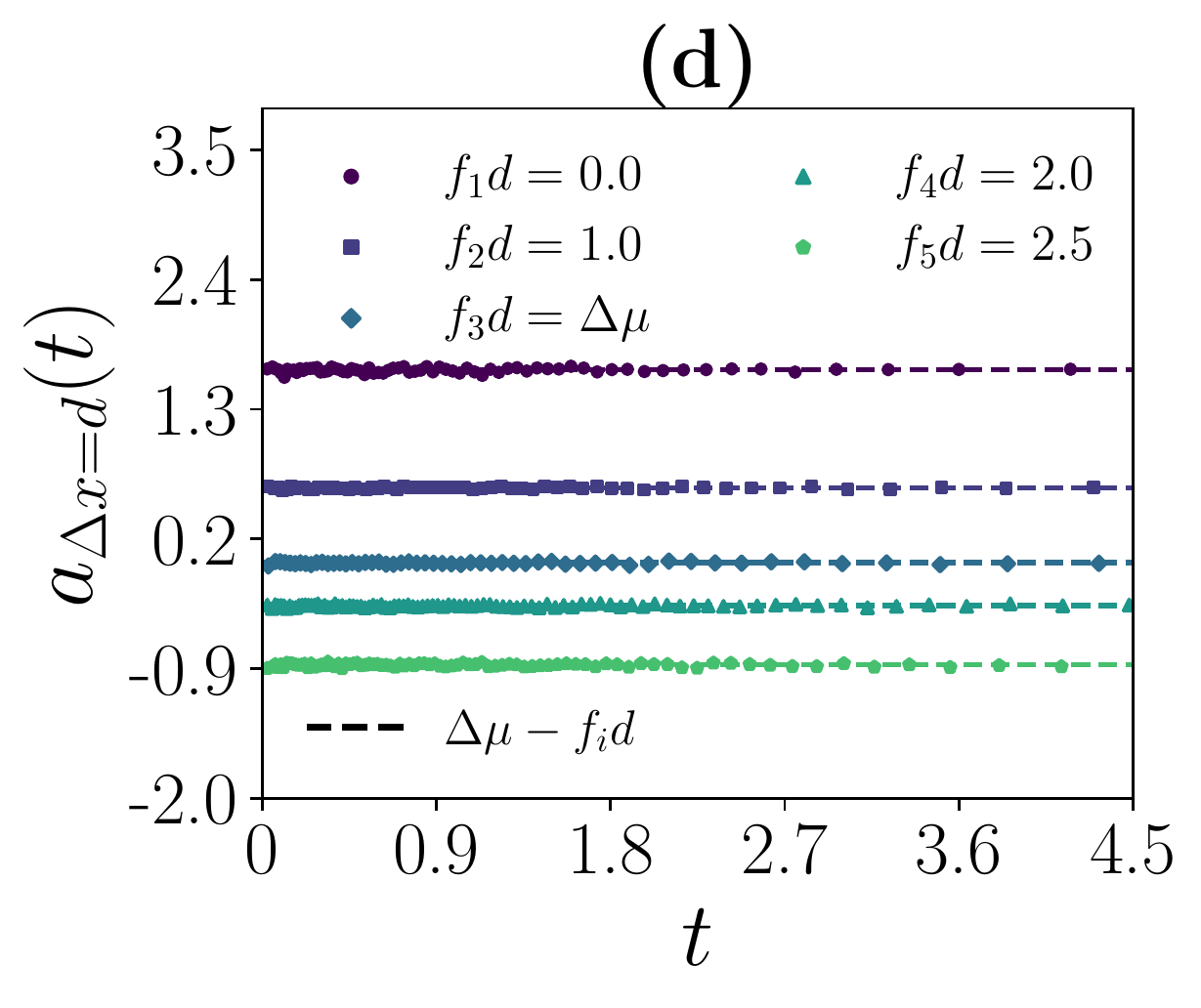}
    \end{subfigure}
    \hfill
    \begin{subfigure}[t]{0.32\textwidth}
      \centering
        \includegraphics[width=\linewidth]{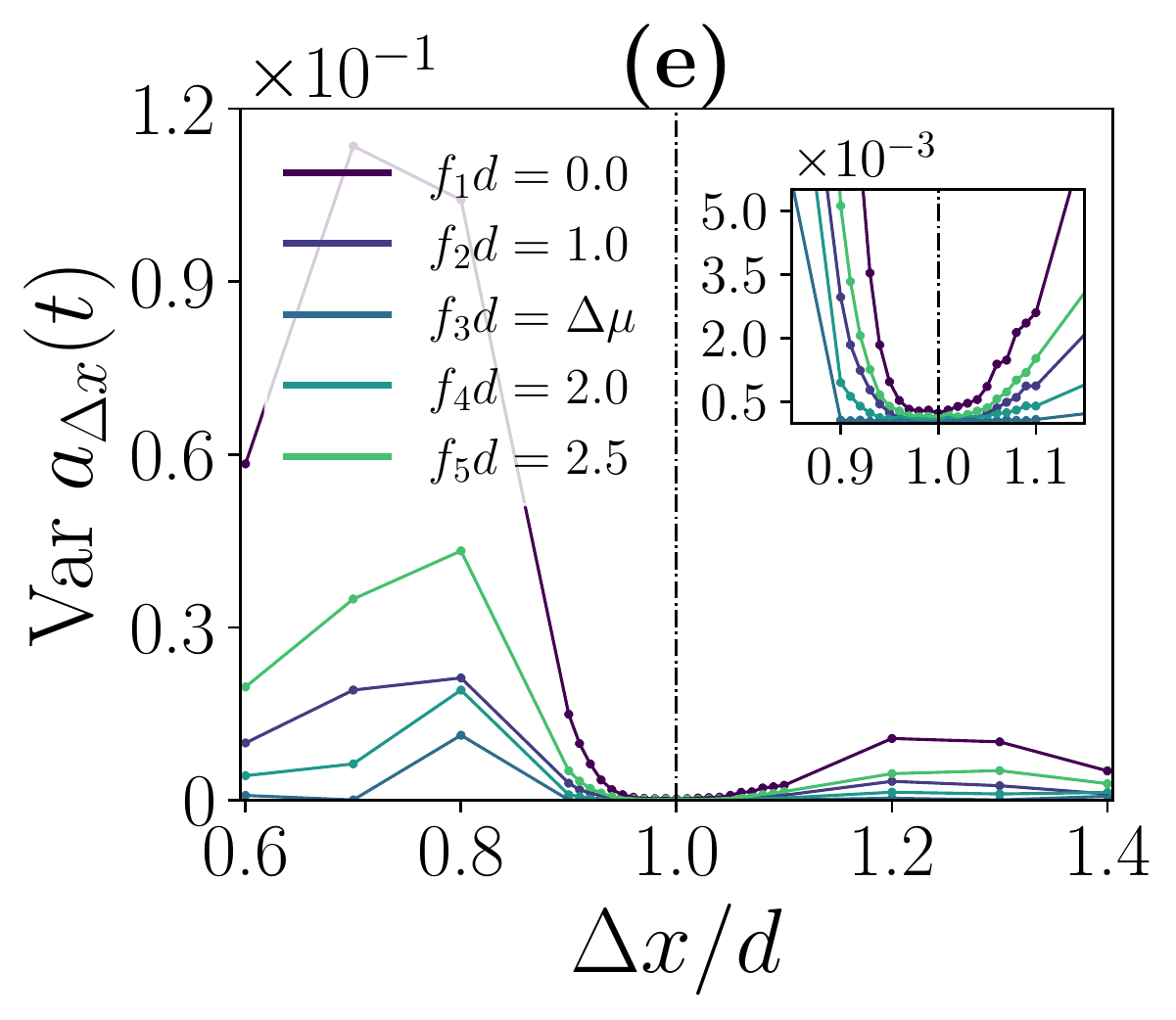}
    \end{subfigure}
    \hfill
    \begin{subfigure}[t]{0.325\textwidth}
      \centering
        \includegraphics[width=\linewidth]{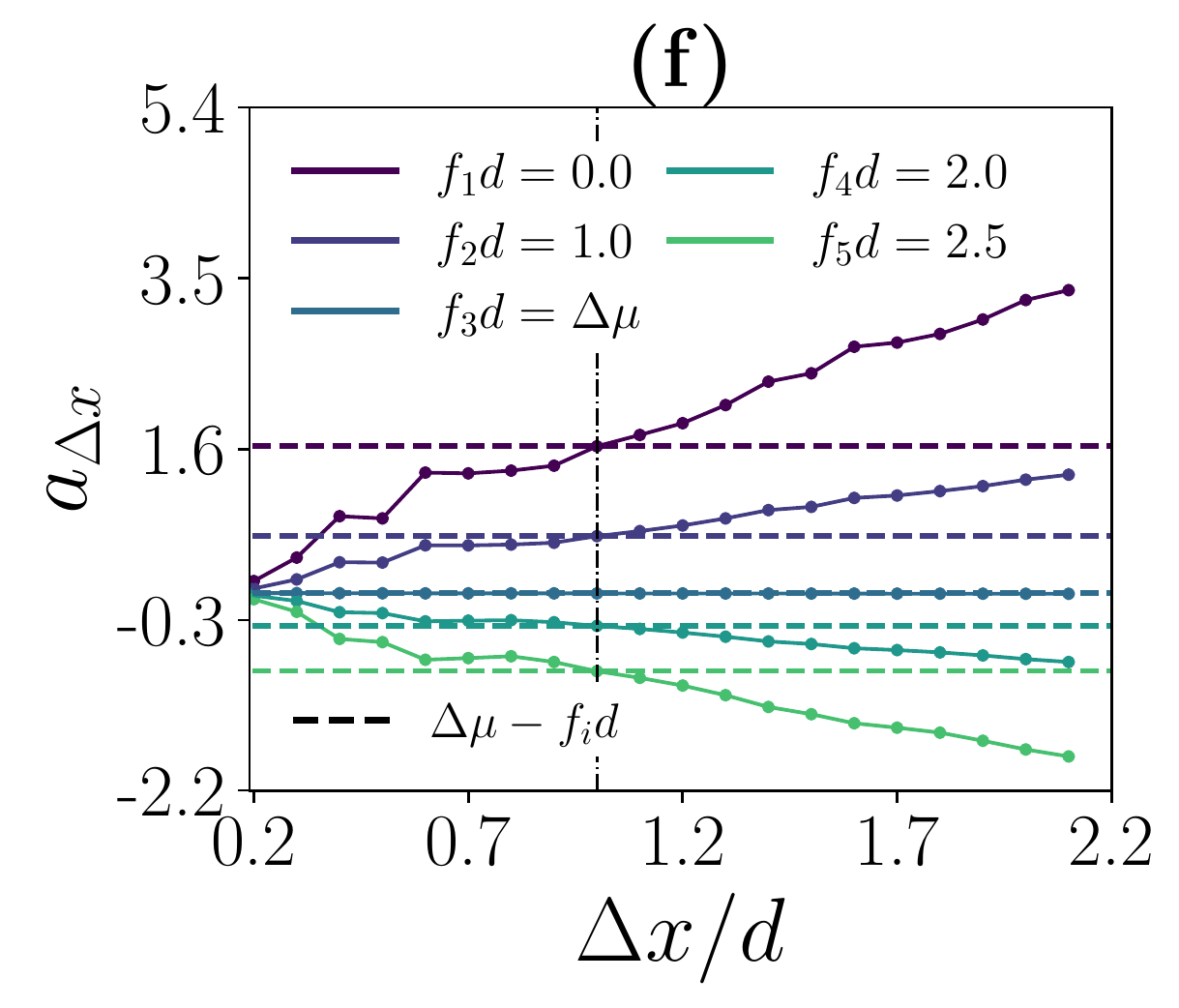}
    \end{subfigure}
    \caption[F1_Normal]{Inference of motor characteristics in the strong coupling regime ($\kappa = 40 d^{-2}$\change{,$\;\Delta\mu = 1.63$}). (a) Exemplary bead trajectory \change{for $fd = 0.0$}. For the chosen coupling strength, single motor steps can be identified along the trajectory. (b), (c) $a_{\Delta x}(t)$ as a function of $t$ for different $\Delta x /d$ \change{for $fd =0.0$}. (d) $a_{\Delta x}(t)$ as a function of $t$ at $\Delta x /d = 1.0$ for different $f$. (e) Variance of $a_{\Delta x}(t)$ as a function of $\Delta x /d$ for different $f$. (f) $a_{\Delta x}$ as a function of $\Delta x /d$ for different $f$. For (d), (e) and (f), the applied force $f_3$ is the stalling force of the motor. In (e) and (f), the black dash-dotted line corresponds to $\Delta x /d = 1.0$.}
\label{Fig:F1_Normal}
\end{figure*}

\section{Application to F1-ATPase}\label{Sec:Hybrid}

\subsection{Simulation method and model parameters}

To study the operational value of the deduced inference quantities, we numerically generate bead trajectories for the hybrid model with parameters from Ref. \cite{zimmermann2012} corresponding to the experimentally realized F1-ATPase assays from Ref. \cite{toyabe2010}. We simulate the dynamics with a Gillespie simulation of the discretized equivalent Fokker-Planck equation \cite{Gillespie_1977,gaspard2007,zimmermann2012}. 

A generated trajectory consists of bead positions and the accompanying time instants. For a sufficient length, counting transitions for a given value of $\Delta x$ results in the corresponding transition probabilities and therefore in $a_{\Delta x}$. For $a_{\Delta x}(t)$, we numerically calculate histograms for the waiting times of the counted transitions. To increase the precision of these histograms, we chose variable bin sizes with a fixed number of samples per bin. Clearly, for larger values of $\Delta x$ and high driving affinities, longer trajectories are needed to deduce meaningful transition statistics. 

Note that describing the F1-ATPase assay with a hybrid model on a one-dimensional line requires mapping a rotational step of $120^{\circ}$ to a motor step of length $d$. The hybrid model does not include the intermediate step of F1-ATPase reported in \cite{yasuda2001,bilyard2013,martin2014}. However, due to the large size of the beads used in experiments, \change{this} intermediate step is not resolved on the level of single trajectories for our choice of parameters \cite{zimmermann2012}. Thus, $\gamma$ sets the noise strength and overall time scale of the bead dynamics. Consequently, $a_{\Delta x}$ and $a_{\Delta x}(t)$ depend on $\gamma$ only marginally. Therefore, we fix the value of $\gamma$ to $\gamma = 0.5 \text{s}\; d^{-2}$ for all simulations. For the chosen set of parameter values, time is measured in seconds. Furthermore, the values of the intrinsic motor quantities $\Theta_+$, $\Theta_-$ and $w_0$ are fixed as the values deduced in Ref. \cite{zimmermann2012}\change{, i.e., $\Theta_+ = 0.1$, $\Theta_- = 0.9$ and $w_0 = 3\cdot 10^7\; \text{M}^{-1}\text{s}^{-1}c_{ATP}^{eq}$}.

For $\kappa$, the coupling strength of the F1-ATPase assay, we distinguish two different regimes. In the experimentally realized coupling regime, which we denote as strong coupling regime, single motor steps are resolved on the bead trajectory level. In the low coupling regime, an identification of single motor steps along the bead trajectory is not possible. Typical bead trajectories for the strong coupling regime and the low coupling regime are shown in Figure~\ref{Fig:F1_Normal} a) and Figure~\ref{Fig:F1_Low} a), respectively. Crucially, we will demonstrate that the operational value of the deduced inference quantities is independent of the coupling regime. In the following two sections, we discuss this operational value for both coupling regimes before rationalizing it in the context of known theoretical results in Section~\ref{Sec:Discussion}.

\subsection{Strong coupling regime}

For strong motor-bead coupling, observable bead transitions are, as illustrated in Figure~\ref{Fig:F1_Normal} a), closely related to the corresponding motor steps. If the chosen $\Delta x$-spacing matches the step size of the motor, i.e., $\Delta x = d$, $a_{\Delta x}(t)$ is constant in time and equal to the driving affinity of the system. In contrast, if we choose $\Delta x$-spacings that do not match the step size of the motor, i.e., $\Delta x \neq d$, $a_{\Delta x}(t)$ is time-dependent with varying shape for different $\Delta x$. This observation is illustrated in Figure~\ref{Fig:F1_Normal} b), c) and d).

The difference between a constant and non-constant $a_{\Delta x}(t)$ can be quantified by calculating the variance of the histogram data that was used to calculate this ratio. As illustrated in Figure~\ref{Fig:F1_Normal} e), the correct step size $\Delta x = d$ is the one for which $\text{Var }a_{\Delta x}(t)$ is minimized. This criterion holds for all chosen $f$ except for the stalling force, i.e., $fd = \Delta\mu$. At stalling, $a_{\Delta x}(t)\approx 0$ for all $\Delta x$ which implies that the variance changes only marginally.  

As illustrated in Figure~\ref{Fig:F1_Normal} f) for different $f$, the driving affinity of the system is equivalently encoded in $a_{\Delta x}$ for $\Delta x = d$ or any integer multiple, i.e.,
\begin{equation}
   a_{\Delta x} = (\Delta\mu - f d)\frac{\Delta x}{d}
   \label{Eq:Hyb_a_PoP}
.\end{equation}
This observation implies that the stalling force $f_\text{stall}$ can directly be recovered from $a_{\Delta x = d}$ for a single value of $f$. Since $a_{\Delta x} = 0$ holds at stalling for all $\Delta x$, we find, using $\Delta\mu = f_\text{stall}d$,
\begin{equation}
    a_{\Delta x = d} = \left(f_\text{stall} - f \right) d
    \label{Eq:Hyb_a_PoP_2}
\end{equation}
for the $a_{\Delta x = d}$ value of a given $f$. 

\subsection{Weak coupling regime}

\begin{figure*}[bt]
    \begin{subfigure}[t]{0.33\textwidth}
      \centering
      \vspace{-4.95cm}
      \hspace{-0.35cm}
        \includegraphics[width=\linewidth]{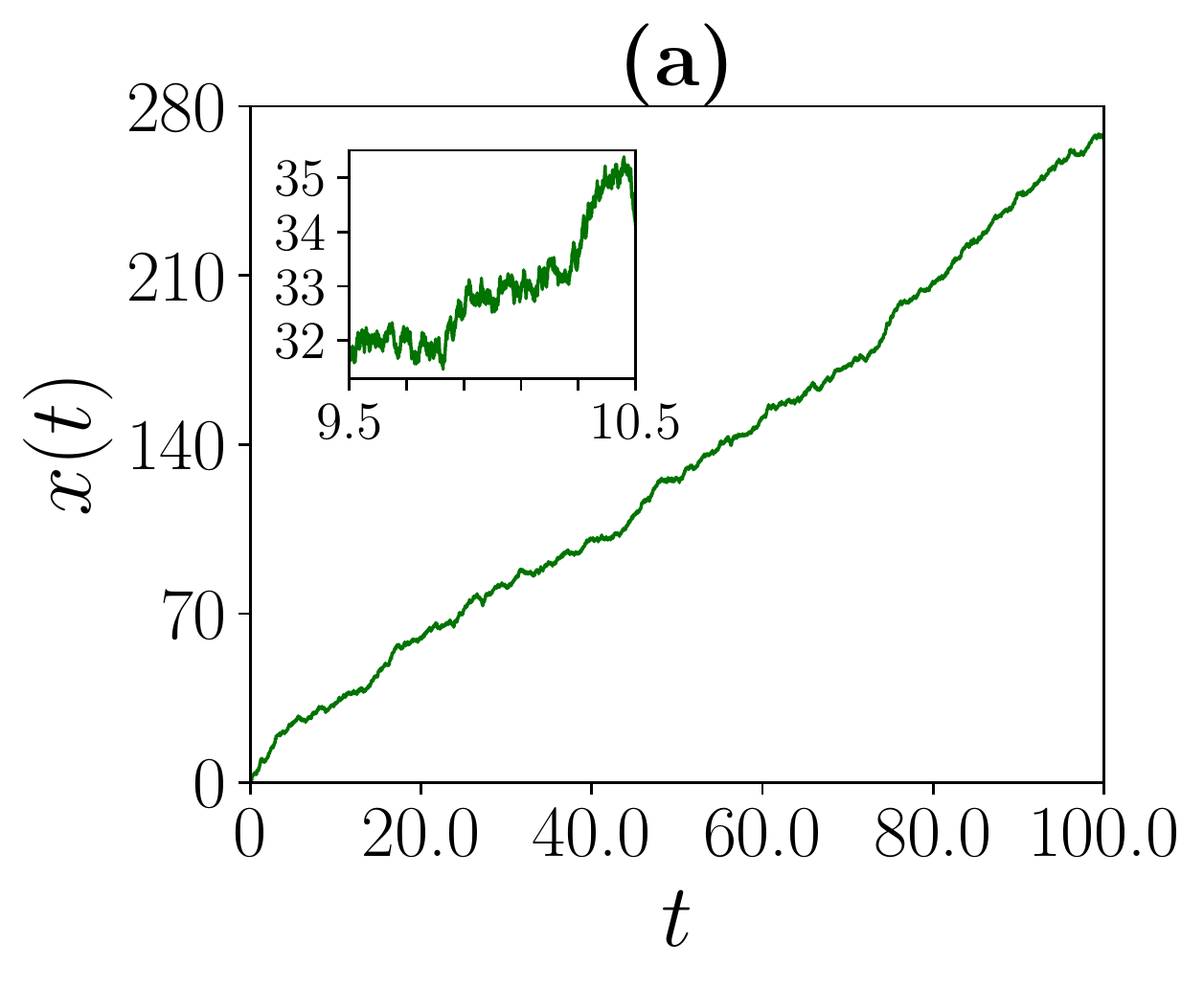}
    \end{subfigure}
    \hfill
    \begin{subfigure}[t]{0.325\textwidth}
      \centering
        \includegraphics[width=\linewidth]{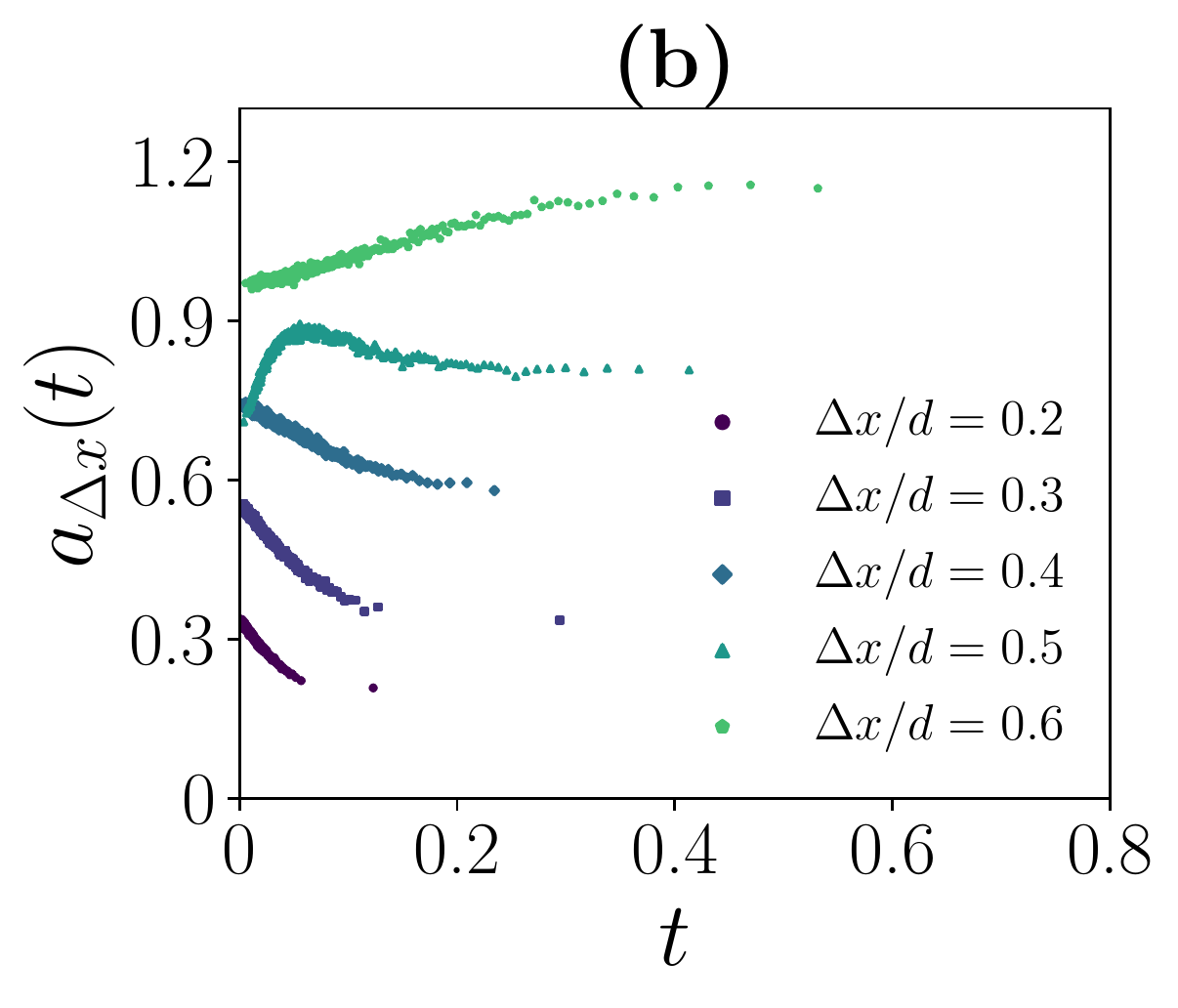}
    \end{subfigure}
    \hfill
    \begin{subfigure}[t]{0.325\textwidth}
      \centering
        \includegraphics[width=\linewidth]{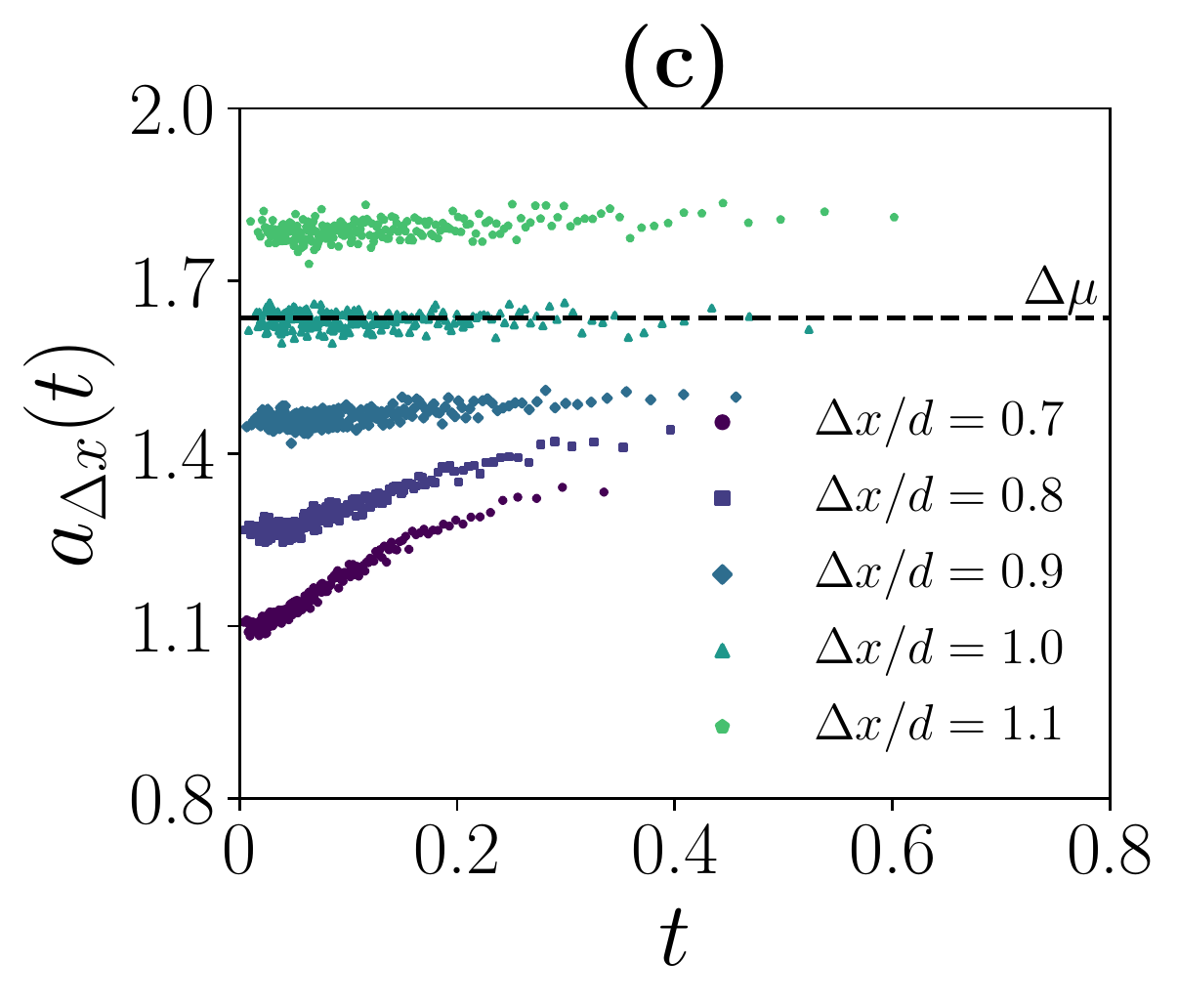}
    \end{subfigure}
    \begin{subfigure}[t]{0.332\textwidth}
      \centering
        \includegraphics[width=\linewidth]{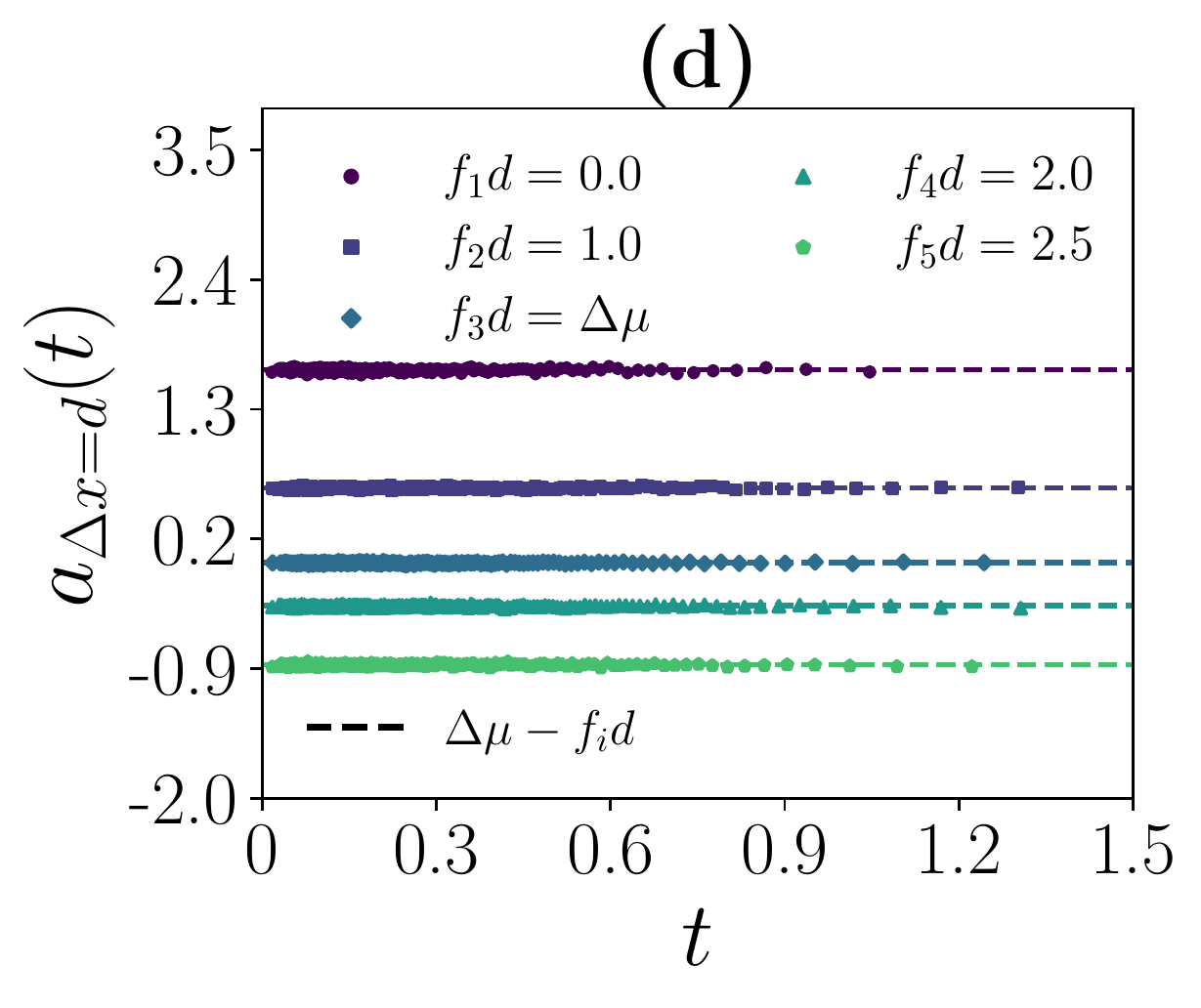}
    \end{subfigure}
    \hfill
    \begin{subfigure}[t]{0.32\textwidth}
      \centering
        \includegraphics[width=\linewidth]{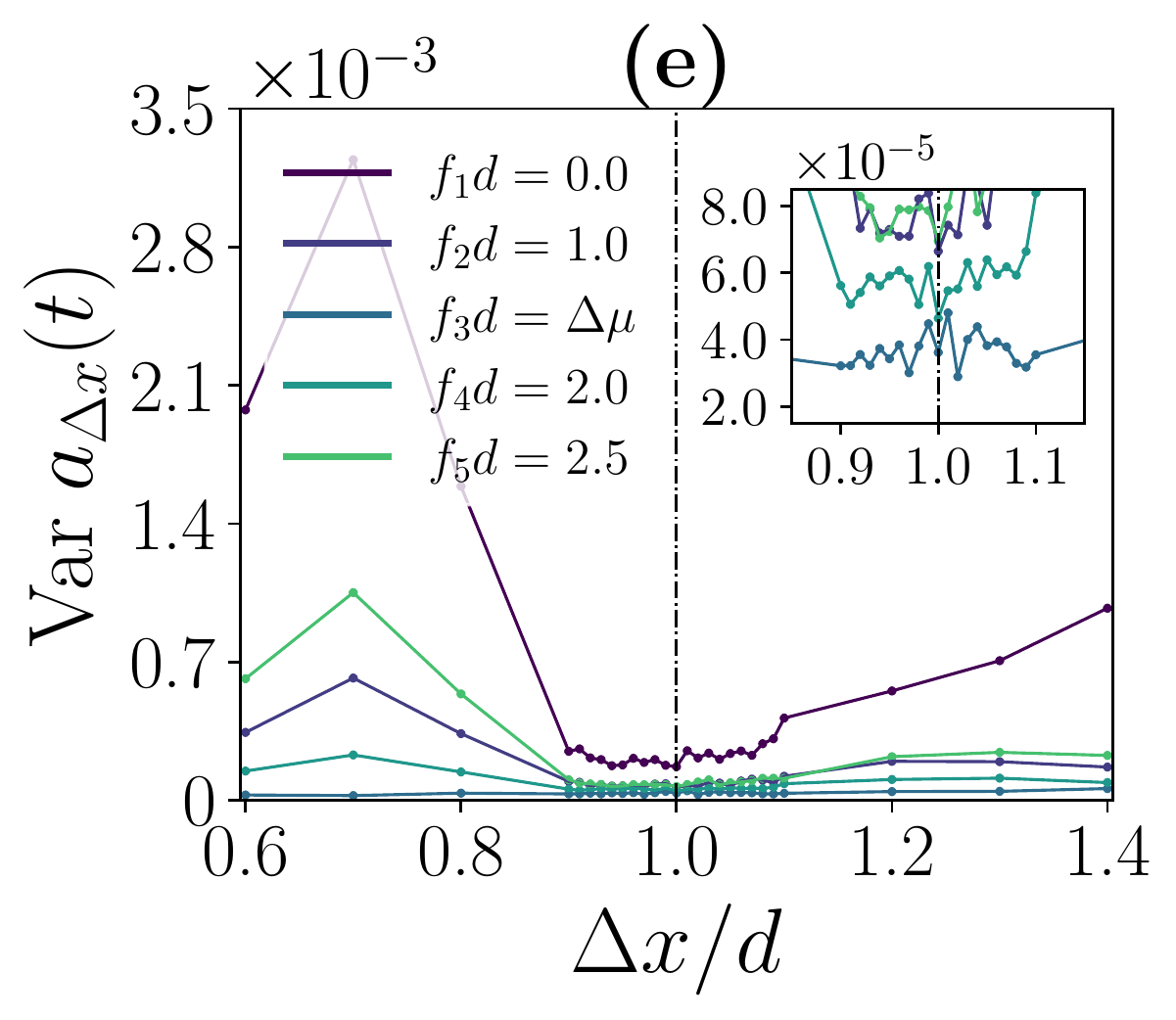}
    \end{subfigure}
    \hfill
    \begin{subfigure}[t]{0.325\textwidth}
      \centering
        \includegraphics[width=\linewidth]{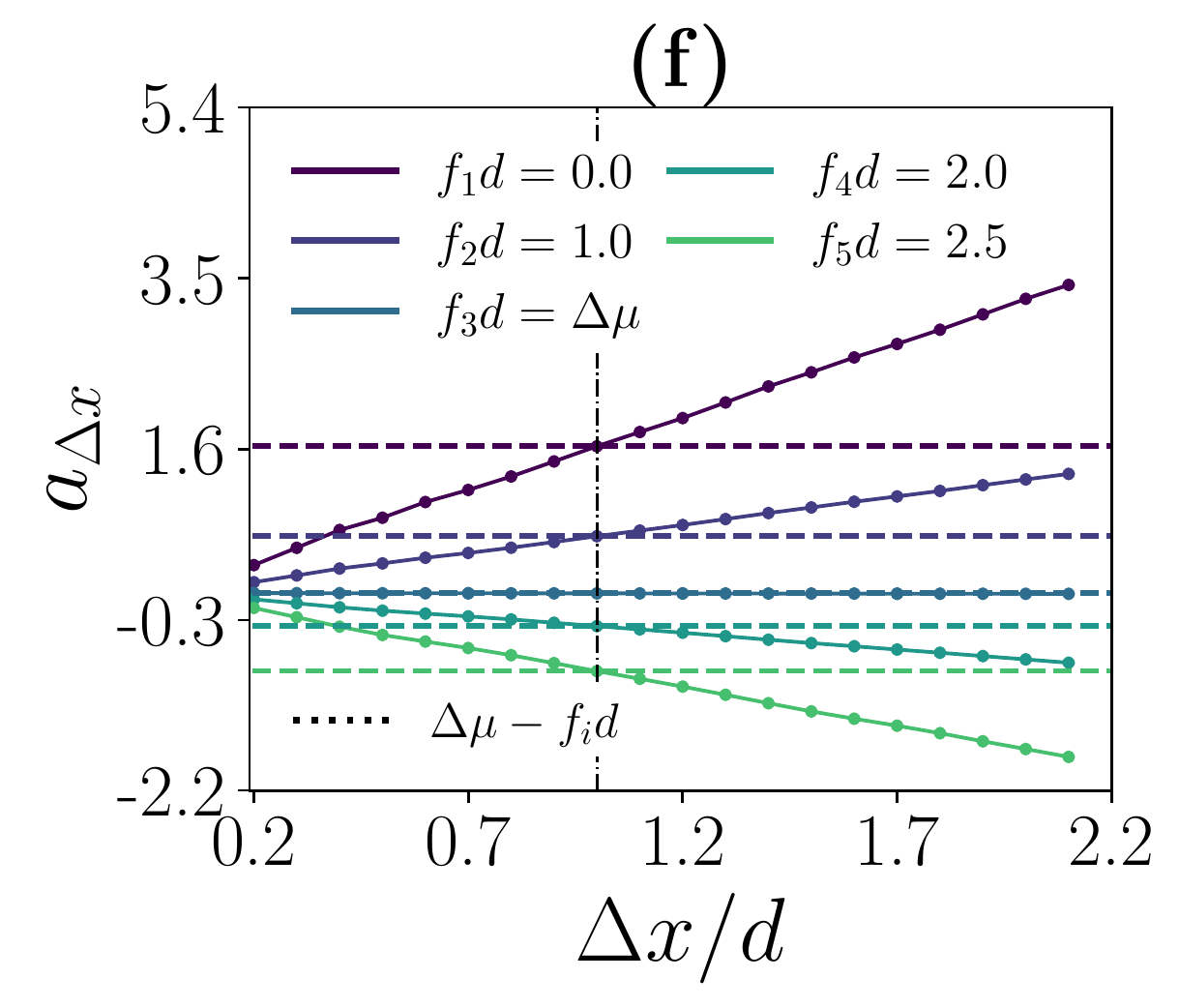}
    \end{subfigure}
    \caption[F1_Low]{Inference of motor characteristics in the regime of weak coupling ($\kappa = 15 d^{-2}$\change{,$\;\Delta\mu = 1.63$}). (a) Exemplary bead trajectory \change{for $fd = 0.0$}. For the chosen coupling strength, an identification of single motor steps along the trajectory is not possible. (b), (c) $a_{\Delta x}(t)$ as a function of $t$ for different $\Delta x /d$ \change{for $fd =0.0$}. (d) $a_{\Delta x}(t)$ as a function of $t$ at $\Delta x /d = 1.0$ for different $f$. (e) Variance of $a_{\Delta x}(t)$ as a function of $\Delta x /d$ for different $f$. (f) $a_{\Delta x}$ as a function of $\Delta x /d$ for different $f$. For (d), (e) and (f), the applied force $f_3$ is the stalling force of the motor. In (e) and (f), the black dash-dotted line corresponds to $\Delta x /d = 1.0$.}
\label{Fig:F1_Low}
\end{figure*}

If motor and bead are weakly coupled, the step size of the motor is, as illustrated in Figure~\ref{Fig:F1_Low} a), not evident from the observation of the bead trajectory. Remarkably, for $\Delta x = d$, $a_{\Delta x}(t)$ is nevertheless constant in time and equal to the driving affinity of the system. Furthermore, as in the strong coupling regime, for $\Delta x \neq d$, $a_{\Delta x}(t)$ is time-dependent with varying shape. This observation is illustrated in Figure~\ref{Fig:F1_Low} b), c) and d).

Similarly to the strong coupling regime, the minimum of $\text{Var }a_{\Delta x}(t)$ is reached at $\Delta x = d$ which yields a quantitative criterion for identifying constant $a_{\Delta x}(t)$. As illustrated in Figure~\ref{Fig:F1_Low} e), this criterion again holds for all chosen $f$ except for the stalling force.
Additionally, Equation~\ref{Eq:Hyb_a_PoP} remains valid in the weak coupling regime, i.e., $a_{\Delta x}$ equals the driving affinity of the system for $\Delta x = d$ or any integer multiple. This observation is illustrated in Figure~\ref{Fig:F1_Low} f) for different $f$. Again, Equation~\ref{Eq:Hyb_a_PoP_2} holds which implies that the stalling force can be recovered from $a_{\Delta x = d}$ for a single $f$.

\section{Discussion}\label{Sec:Discussion}

\subsection{Relation to established inference tools}\label{sec:theory}

The logarithmic ratios of waiting time distributions and transition probabilities, i.e, $a_{\Delta x}(t)$ as defined by Equation~\ref{Eq:a_deltax_t} and $a_{\Delta x}$ defined by Equation~\ref{Eq:a_deltax}, are antisymmetric under time-reversal. Furthermore, these observables show conceptual similarities to transition paths \cite{berezhkovskii2020,berezhkovskii2021,hartich2021,makarov2022,godec2023} and observed transitions in partially accessible Markov networks \cite{PRX,harunari2022}, which implies that they are suited to detect irreversibility. More precisely, for partially accessible Markov networks, it was shown in Ref. \cite{PRX} using cycle-based fluctuation theorem arguments that the logarithmic ratio of conditioned transition probabilities
\begin{equation}
    a = \ln \frac{p(+|+)}{p(-|-)}
    ,\label{Eq:a_NoT}
\end{equation}
which corresponds to $a_{\Delta x}$ in our framework, encodes information about thermodynamic properties and topological characteristics of the full Markov network. For example, if this network contains a single cycle, in which the forward and backward direction of a particular transition are registered as ''$+$'' and ''$-$'', respectively, $a$ coincides with the driving affinity of the cycle. This result holds, in principle, for transitions between \change{$(y,x)$} and \change{$(y \pm d,x \pm d)$} in the hybrid model as well, because the joint dynamics of \change{$(y(t),x(t))$} is Markovian. Furthermore, translational invariance ensures that \change{$(y,x)$} and \change{$(y \pm d,x \pm d)$} can indeed be treated as the same state, which implies that the hybrid model is \emph{de facto} unicyclic. Therefore, $a_{\Delta x }$ recovers for $\Delta x = d$ the driving affinity of the motor in both coupling regimes. 

However, in practice, our observation only registers bead transitions between $x$ and $x \pm d$ without knowledge about the motor position \change{$y$}, which at first sight does not allow for a Markovian description. Nevertheless, we can utilize waiting time distributions to quantify the ''non-Markovianity'' of the observed dynamics. In particular, the time-dependent counterpart of Equation~\ref{Eq:a_NoT}, 
\begin{equation}
    a(t) = \ln \frac{\psi_{+ \to +}(t)}{\psi_{- \to -}(t)}
    \label{Eq:a_T}
,\end{equation}
which corresponds to $a_{\Delta x}(t)$ in our framework, remains constant in one-dimensional Langevin dynamics \cite{berezhkovskii2006} and unicyclic Markov networks \cite{PRX}, even out of equilibrium. In a fairly general setup, a violation of this symmetry property for transitions is either a consequence of the non-Markovian character of the dynamics or hints at the presence of hidden cycles \cite{berezhkovskii2019,PRL,godec2023}. Operationally, for the hybrid model, the preceding theoretical discussion yields a criterion for the applicability of a Markovian approximation. Notably, this approach is fundamentally different from comparing and fitting observed waiting time distributions to the characteristic exponential distributions for discrete Markov states, which is not applicable to transitions in a continuous framework.   

\subsection{The crucial role of waiting time distributions}

For a one-dimensional Markovian random walk on a discrete lattice with step size $\Delta x$, the actual motor dynamics without an attached bead, we have $a_{\Delta x}(t) = \text{const.}$ for any choice of parameters. Deviations from this random walk remain small as long as the discrete description of the bead position $x(t)$ essentially captures the dynamics of \change{$(y(t),x(t))$}, i.e., the dynamics of both motor and bead. This condition holds if \change{$p(y(t) - x(t)|x(t))$}, the uncertainty in \change{$y(t) - x(t)$} given the bead position $x(t)$, remains sharply peaked around $0$. Stated differently, for the correct step size, the Markovian random walk preserves essential features of the dynamics, as forward or backward steps of the motor match the corresponding observable transitions of the bead. Clearly, in the strong coupling regime, the bead dynamics fulfill this requirement. Notably, in the weak coupling regime, this requirement is analogously fulfilled, although bead dynamics and motor movement are apparently uncorrelated. 

The situation is different for $\Delta x$-spacings that differ from the correct step size of the motor. Due to the discrepancy between the length $d$ of the motor steps and the chosen length $\Delta x$ for detected transitions, there is in general no correspondence between these, even if the bead follows the motor closely. Thus, the behavior of $a_{\Delta x}(t)$ for small and large $t$ differs, which generally leads to qualitatively different and more complex shapes of $a_{\Delta x}(t)$ and hence to a higher value of $\text{Var }a_{\Delta x}(t)$ compared to the variance of a constant $a_{\Delta x}(t)$. 

\begin{figure}[bt]
	\begin{center}
    \begin{subfigure}[t]{0.55\textwidth}
      \centering
        \includegraphics[width=\linewidth]{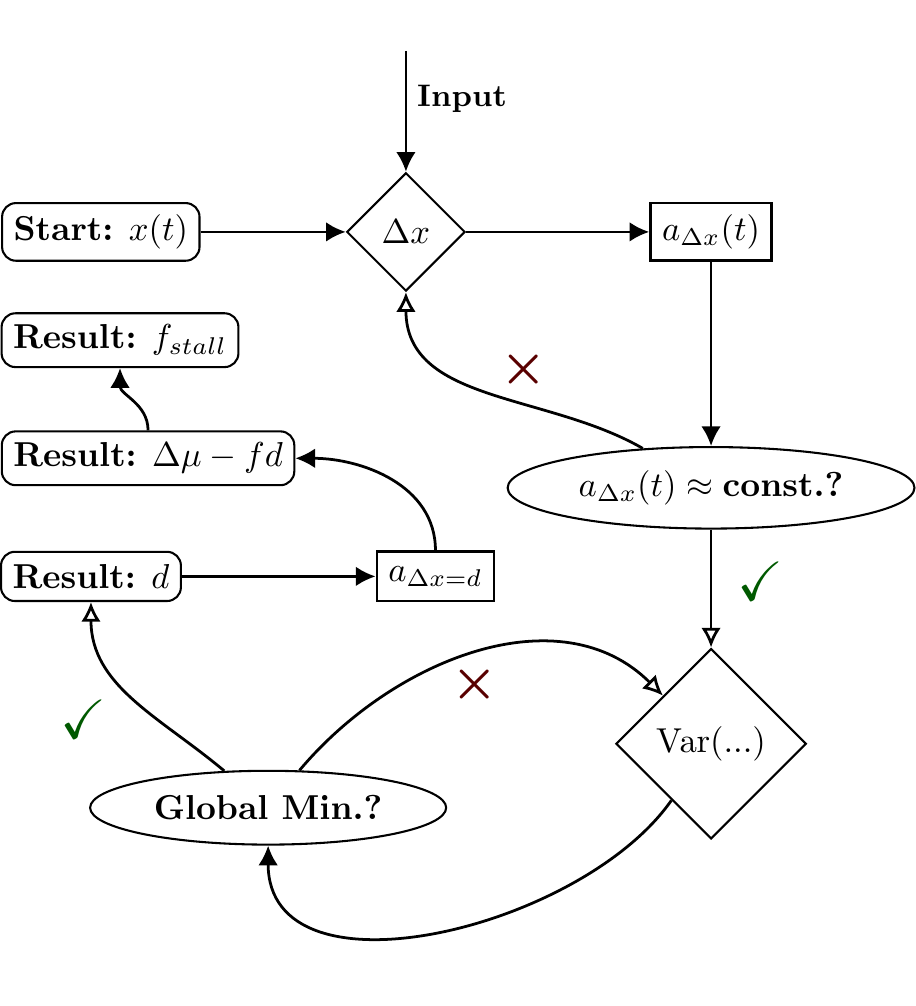}
    \end{subfigure}
    \end{center}
    \caption[Flowchart]{Illustration of the introduced inference procedure. Starting from an observed bead trajectory $x(t)$, we calculate $a_{\Delta x}(t)$ for different $\Delta x$. Around values of $\Delta x$ with approximately constant $a_{\Delta x}(t)$, we identify the minimal value of the variance of $a_{\Delta x}(t)$. The resulting $\Delta x$ corresponds to the motor step width $d$. For $\Delta x = d$, the non-time-resolved $a_{\Delta x}$ is equal to $\Delta\mu - fd$ which yields $f_\text{stall}$.}
\label{Fig:Flow}
\end{figure}

Thus, for an observed bead trajectory, evaluating the time-dependence of $a_{\Delta x}(t)$ for different values of $\Delta x$ allows us to find the values of $\Delta x$ for which a Markovian approximation can be applied and therefore, the inference results for partially accessible Markov networks or, more generally, waiting time distributions between particular observed events hold. Combined with minimizing $\text{Var }a_{\Delta x}(t)$, this approach recovers the step size $d$ of the motor. In a second step, the driving affinity can be inferred from $a_{\Delta x}$ at $\Delta x = d$. Additionally, as proven in Ref. \cite{PRX}, $a_{\Delta x}$ recovers the stalling force. The operational details and the single steps of the suggested inference procedure are summarized in Figure~\ref{Fig:Flow}.

We point out that this inference procedure is fundamentally different to prescribing a Markov model directly. While a Markov model, if applicable, allows extracting the driving affinity too, our approach includes a criterion whether the Markov approximation is viable or not. Furthermore, a Markov model has to include assumptions about motor characteristic, for example the correct step size $\Delta x = d$ of the bead. Crucially, within our non-invasive approach, the step size is not required as an input for but can rather be inferred from the observation of bead trajectories alone.  

\subsection{Alternative observables}

The detection of a wrong $\Delta x$-spacing can utilize additional information in a similar fashion to waiting times. For example, we can consider the unconditioned logarithmic ratio
\begin{equation}
    \tilde{a}_{\Delta x} = \ln \frac{p(+)}{p(-)} =  \ln \frac{n_{\Delta x}^+}{n_{\Delta x}^-}
\end{equation}
and its time-dependent counterpart
\begin{equation}
    \tilde{a}_{\Delta x}(t) = \ln \frac{\psi^{\Delta x}_{+ \to +}(t) + \psi^{\Delta x}_{- \to +}(t)}{\psi^{\Delta x}_{+ \to -}(t) + \psi^{\Delta x}_{- \to -}(t)}
\end{equation}
as an alternative to Equation~\ref{Eq:a_deltax} and Equation~\ref{Eq:a_deltax_t}, respectively. For $\Delta x = d$ we expect $\tilde{a}_{d}(t) = \text{const.} = a_{d}(t)$, since the dynamics is essentially a Markov random walk and therefore memoryless, i.e., independent of conditioning on past events. Turning the argument around, if a ratio of waiting time distributions like $\tilde{a}_{\Delta x}$ is sensitive to selecting particular times, conditions, etc., this hints at hidden memory effects like, \emph{e.g.}, discrepancies between the positions of motor and bead for the hybrid model. For example, $a_{\Delta x} \neq \tilde{a}_{\Delta x}$ for a particular value of $\Delta x$ is an evidence against $\Delta x$ as the correct step size and therefore an evidence against the applicability of a Markovian approximation for this value of $\Delta x$ .

\section{Conclusions}\label{Sec:Conclusion}

In this paper, we have translated principles of thermodynamic inference based on observable transitions to a generic hybrid model for motor-bead assays. Based on a coarse-grained description of observable bead dynamics focusing on transitions, we have introduced a procedure for inferring the exact step width and the total driving affinity, i.e., the stalling force, of the non-observable motor. To illustrate its operational significance, we have inferred the aforementioned motor characteristics from simulated bead trajectories of the model for parameters that describe an experimentally realized F1-ATPase assay in different coupling regimes.  

\change{Generalizing the minimal hybrid model in the spirit of Ref. \cite{zimmermann2015} can provide a starting point for subsequent studies. As shown there, such a generalization reproduces the correct dynamical behavior for elementary molecular motors like the F1-ATPase and more complicated motor proteins with internal cycles. In fact, in several biological and synthetic motors with a more complicated mechanochemical structure, for example kinesin-1, such internal cycles are possible, e.g., in the form of idle cycles, which lead to additional dissipation but remain unobserved in mechanical steps \cite{ariga2018}. Hence, a complete motor-bead assay model then has to include internal cycles within the motor, which results in multicyclic internal motor dynamics. From a technical point of view, considering these more general hybrid models would allow one to incorporate more advanced inference results for partially accessible multicyclic Markov networks into the inference procedure. In such a modified procedure, we might be able to infer further motor characteristics. For example, taking the short-time limit yields topological information about short cycles and the driving affinity of the shortest cycle \cite{PRX}. From a conceptual point of view, this kind of additional information could potentially even aid in unraveling the mechanochemical structure of more complex motor proteins.} 

%Generalizing the minimal hybrid model in the spirit of \cite{zimmermann2015} provides a starting point for subsequent studies. While the hybrid model reproduces correct dynamical behavior for elementary molecular motors like the F1-ATPase, it is tailor-made for motors without internal cycles. However, in several biological and synthetic motors with a more complicated mechanochemical structure, such internal cycles are possible, e.g., in the form of idle cycles, which cannot be observed in mechanical steps. Hence, motor-bead assay models like the hybrid model have to include internal cycles within the motor, which invites applying more advanced inference results for partially accessible multicyclic Markov networks.

A different direction of future research can be pursued by applying the inference scheme to different molecular motor models, in particular ratchet models \cite{juelicher1997,aithaddou2003,kolomeisky_book}. Despite the fundamental difference to hybrid models on a conceptual level, ratchet models produce superficially similar bead trajectories. On the one hand, the underlying descriptions could become distinguishable in the waiting time statistics of appropriate transition events. On the other hand, a successful application of the methods here to ratchet models would establish a tool for inference beyond a particular model class.

Similarly, future work could aim at applying the inference methods to compare predictions of particular models with experimental data. As long as the statistics extracted from experiments suffice to extract single transitions or even waiting time distributions, predictions of particular models can be falsified from non-invasive measurements alone. Thus, our operationally accessible results offer a novel approach to study the range of applicability of motor-bead assay models with varying degrees of sophistication, from discrete Markov random walks to models that contain motor states and continuous degrees of freedom.

\section*{Acknowledgements}
We thank Julius Degünther for many valuable discussions.

%\bibliography{referencesNeu}
%\bibliographystyle{apsrev4-2}

\input{output.bbl}

\end{document}